\documentclass[11pt]{cernrep}
\usepackage{graphicx}
\usepackage{cite,./mcite}
\usepackage{axodraw}
\begin{document}
\title{
\begin{flushright}
{\normalsize\mdseries 
\textrm{LU TP 08--17}\\
\textrm{MCnet/08/09}\\
\textrm{September 2008}\\[5mm]}
\end{flushright}
PYTHIA 8 Status Report%
\footnote{\hspace*{2mm}to appear in the proceedings of the 
HERA and the LHC workshop, 26--30 May 2008, CERN}}
\author{Torbj\"orn Sj\"ostrand}
\institute{Department of Theoretical Physics, Lund University}
\maketitle
\begin{abstract}
\textsc{Pythia}~8, the C++ rewrite of the commonly-used \textsc{Pythia}
event generator, is now available in a first full-fledged version 8.1. 
The older \textsc{Pythia}~6.4 generator in Fortran 77 is still maintained, 
for now, but users are strongly recommended to try out and move to the 
new version as soon as feasible.
\end{abstract}

\section{Introduction}

The ``Lund Monte Carlo'' family of event generators started in 1978
with the \textsc{Jetset} program. \textsc{Pythia} was begun a few years 
later, and the two eventually were joined under the \textsc{Pythia} label. 
Over the last 25 years the \textsc{Pythia}/\textsc{Jetset} program has been 
widely used to help understand the physics of high-energy collisions.

The program was from the onset written in Fortran 77, up to the current 
version 6.4 \cite{Sjostrand:2006za}. However, following the move of the 
experimental community to C++, a corresponding restart and rewrite was 
made for \textsc{Pythia} in 2004 -- 2007, with most aspects cleaned up 
and modernized.

The first production quality release, \textsc{Pythia}~8.100, appeared 
towards the end of 2007 \cite{Sjostrand:2007gs}. It was paced to arrive 
in time for LHC and therefore does not yet cover some physics topics. 
It has not yet caught on in the LHC experimental collaborations, however, 
and thus the older Fortran code is still maintained, even if at a reduced 
level. 

\section{Physics summary}

Here follows a brief summary of the key physics aspects of 
\textsc{Pythia}~8.1, by topic.

{\bf Hard processes:} The built-in library contains many leading-order
processes, for the Standard Model almost all $2 \to 1$ and $2 \to 2$ ones
and a few $2 \to 3$, beyond it a sprinkling of different processes, but not
yet Supersymmetry or Technicolor. Parton-level events can also be input from
external matrix-element-based generators, e.g. using Les Houches Event Files
\cite{Alwall:2006yp}. Also runtime interfaces are possible, and one such is 
provided to \textsc{Pythia}~6.4 for the generation of legacy processes. 
Resonance decays are included, often but not always with full angular 
correlations.

{\bf Parton showers:} Transverse-momentum-ordered showers are used both for
initial- and final-state radiation, the former based on backwards evolution.
Implemented branchings are $q \to q g$, $g \to g g$, $g \to q \overline{q}$,
$f \to f \gamma$ ($f$ is a quark or lepton) and $\gamma \to f \overline{f}$.
Recoils are handled in a dipole-style approach, but emissions are still 
associated with one emitting parton. Many processes include matching to 
matrix elements for the first (= hardest) emission; this especially 
concerns gluon emission in resonance decays.

{\bf Underlying events and minimum-bias events:} \textsc{Pythia} implements 
a formalism with multiple parton--parton interactions, based on the standard 
QCD matrix elements for $2 \to 2$ processes, dampened in the $p_{\perp} \to 0$ 
limit. The collision rate is impact-parameter-dependent, and collisions are 
ordered in decreasing $p_{\perp}$. Multiple interactions (MI) are therefore 
combined with initial- and final-state radiation (ISR and FSR) in one common 
sequence of decreasing transverse momenta 
$p_{\perp 1} > p_{\perp 2} > p_{\perp 3} \ldots$,
\begin{eqnarray*}
\left. \frac{\mathrm{d} \mathcal{P}}{\mathrm{d} p_{\perp}} 
\right|_{p_{\perp} = p_{\perp i}} & = &
\left(
\frac{\mathrm{d} \mathcal{P}_{\mathrm{MI}}}{\mathrm{d} p_{\perp}} + 
\sum \frac{\mathrm{d} \mathcal{P}_{\mathrm{ISR}}}{\mathrm{d} p_{\perp}} +
\sum \frac{\mathrm{d} \mathcal{P}_{\mathrm{FSR}}}{\mathrm{d} p_{\perp}}
\right) \\
& \times & \exp \left( - \int_{p_{\perp}}^{p_{\perp i-1}} \left(
\frac{\mathrm{d} \mathcal{P}_{\mathrm{MI}}}{\mathrm{d} p_{\perp}'} + 
\sum \frac{\mathrm{d} \mathcal{P}_{\mathrm{ISR}}}{\mathrm{d} p_{\perp}'} +
\sum \frac{\mathrm{d} \mathcal{P}_{\mathrm{FSR}}}{\mathrm{d} p_{\perp}'}
\right) \mathrm{d} p_{\perp}' \right) ~,
\end{eqnarray*}
using the ``winner takes all'' Monte Carlo strategy. This leads to a 
competition, in particular between MI and ISR, for beam momentum. The
beam remnants are colour-connected to the interacting subsystems, with 
a detailed modelling of the flavour and momentum structure, also for
the parton densities to be used at each successive step. The framework also
contains a model for colour reconnection, likely the least well understood 
aspect of this physics area, and therefore one that may require further 
development. 

{\bf Hadronization:} The Lund model for string fragmentation is used to
describe the transition from coloured partons to colour singlet hadrons.
Subsequent hadronic decays are usually described isotropic in phase space,
but in some cases matrix-element information is inserted. It is also 
possible to link to external decay packages, e.g. for $\tau$ or $B$ decays.
A model for Bose--Einstein effects is included, but is off by default.

\section{Program evolution}

The above physics description largely also applies to \textsc{Pythia}~6.4. 
There are some differences to be noted, however.

Many old features have been definitely removed. Most notably this concerns
the framework for independent fragmentation (a strawman alternative to string 
fragmentation) and the older mass-ordered showers (that still are in use
in many collaborations, but do not fit so well with the new interleaved
MI/ISR/FSR description). 

Features that have been omitted so far, but should appear when time permits,
include $e p$, $\gamma p$ and $\gamma \gamma$ beam configurations and a set 
of SUSY and Technicolor processes. 

New features, relative to \textsc{Pythia}~6.4 include 
\begin{itemize}
\item the interleaved MI/ISR/FSR evolution (6.4 only interleaved MI and ISR),
\item a richer mix of underlying-event processes, no longer only QCD jets
but also prompt photons, low-mass lepton pairs and $J/\psi$,
\item possibility to select two hard processes in an event,
\item possibility to use one PDF set for the hard process and another for 
MI/ISR, and
\item updated decay data.
\end{itemize}

Major plans for the future include a new model for rescattering processes 
in the MI machinery, and new facilities to include 
matrix-element-to-parton-shower matching.   

In addition minor improvements are introduced with each new subversion.
Between the original 8.100 and the current 8.108 the list includes
\begin{itemize}
\item possibility to have acollinear beams, beam momentum spread and
beam vertex spread,
\item updated interfaces to several external packages,
\item improved possibility to run several \texttt{Pythia} instances
simultaneously, 
\item code modifications to compile under gcc 4.3.0 with the 
\texttt{-Wshadow} option, and
\item some minor bug fixes.
\end{itemize}

\begin{figure}[t]
\begin{picture}(430,370)(-215,10)
\GBox(-215,350)(215,380){0.9}
\Text(0,365)[]{The User ($\approx$ Main Program)}
\GBox(-215,300)(215,330){0.9}
\Text(0,315)[]{\texttt{Pythia}}
\GBox(-215,250)(-170,280){0.9}
\Text(-192.5,265)[]{\texttt{Info}}
\GBox(-130,250)(-20,280){0.9}
\Text(-75,265)[]{\texttt{Event~~process}}
\GBox(20,250)(215,280){0.9}
\Text(105,265)[]{\texttt{Event~~event}}
\GBox(-215,110)(-85,230){0.9}\Line(-215,200)(-85,200)
\Text(-150,215)[]{\texttt{ProcessLevel}}
\Text(-150,185)[]{\texttt{ProcessContainer}}
\Text(-150,165)[]{\texttt{PhaseSpace}}
\Text(-150,145)[]{\texttt{LHAup}}
\Text(-150,125)[]{\texttt{ResonanceDecays}}
\GBox(-67,110)(67,230){0.9}\Line(-67,200)(67,200)
\Text(0,215)[]{\texttt{PartonLevel}}
\Text(0,185)[]{\texttt{TimeShower}}
\Text(0,165)[]{\texttt{SpaceShower}}
\Text(0,145)[]{\texttt{MultipleInteractions}}
\Text(0,125)[]{\texttt{BeamRemnants}}
\GBox(85,110)(215,230){0.9}\Line(85,200)(215,200)
\Text(150,215)[]{\texttt{HadronLevel}}
\Text(150,185)[]{\texttt{StringFragmentation}}
\Text(150,165)[]{\texttt{MiniStringFrag\ldots}}
\Text(150,145)[]{\texttt{ParticleDecays}}
\Text(150,125)[]{\texttt{BoseEinstein}}
\GBox(-130,60)(-20,90){0.9}
\Text(-75,75)[]{\texttt{BeamParticle}}
\GBox(20,60)(200,90){0.9}
\Text(110,75)[]{\texttt{SigmaProcess, SigmaTotal}}
\GBox(-215,10)(215,40){0.9}
\Text(0,25)[]{\texttt{Vec4, Rndm, Hist, Settings, %
ParticleDataTable, ResonanceWidths}}
\SetWidth{2}
\LongArrow(0,350)(0,332)
\LongArrow(-150,300)(-150,232)
\LongArrow(0,300)(0,232)
\Line(150,300)(150,280)
\DashLine(150,280)(150,250){4}
\LongArrow(150,250)(150,232)
\SetWidth{1}
\LongArrow(-192.5,230)(-192.5,248)
\Line(-192.5,280)(-192.5,300)
\DashLine(-192.5,300)(-192.5,330){4}
\LongArrow(-192.5,330)(-192.5,348)
\LongArrow(-107.5,230)(-107.5,248)
\LongArrow(-42.5,250)(-42.5,232)
\Line(-42.5,280)(-42.5,300)
\DashLine(-42.5,300)(-42.5,330){4}
\LongArrow(-42.5,330)(-42.5,348)
\LongArrow(42.5,230)(42.5,248)
\LongArrow(107.5,250)(107.5,232)
\LongArrow(182.5,230)(182.5,248)
\Line(182.5,280)(182.5,300)
\DashLine(182.5,300)(182.5,330){4}
\LongArrow(182.5,330)(182.5,348)
\LongArrow(-107.5,100)(-107.5,108)
\LongArrow(-107.5,100)(-107.5,92)
\LongArrow(-42.5,100)(-42.5,108)
\LongArrow(-42.5,100)(-42.5,92)
\LongArrow(42.5,100)(42.5,108)
\LongArrow(42.5,100)(42.5,92)
\LongArrow(-160,50)(-160,108)
\Line(-160,50)(0,50)
\LongArrow(0,50)(19,59)
\end{picture}
\caption{The relationship between the main classes in 
\textsc{Pythia}~8. The thick arrows show the flow of commands
to carry out different physics tasks, whereas the thinner show
the flow of information between the tasks. The bottom box 
contains common utilities that may be used anywhere. Obviously 
the picture is strongly simplified.
\label{fig:generatorstructure}}
\hrulefill  
\end{figure}
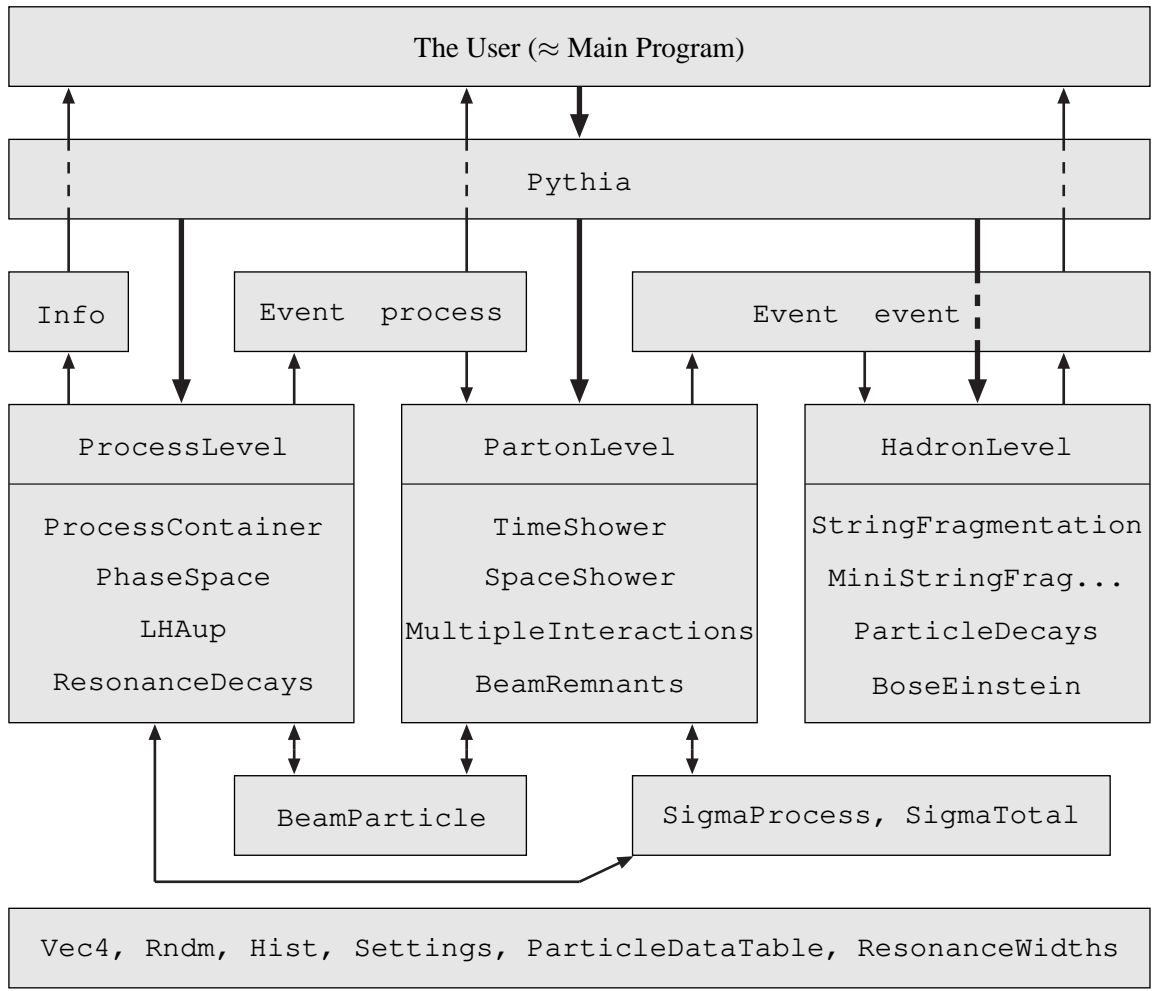

\section{Program structure}
 
The structure of the \textsc{Pythia}~8 generator is illustrated in 
Fig.~\ref{fig:generatorstructure}. The main class for all user 
interaction is called \texttt{Pythia}. It calls on the three classes 
\begin{itemize}
\item \texttt{ProcessLevel}, for the generation of the hard process,
by sampling of built-in matrix elements or input from an external program,
\item \texttt{PartonLevel}, for the additional partonic activity by
MI, ISR, FSR and beam remnants, and
\item \texttt{HadronLevel}, for the transition from partons to hadrons
and the subsequent decays.
\end{itemize}
Each of these, in their turn, call on further classes that perform the 
separate kinds of physics tasks.

Information is flowing between the different program elements in
various ways, the most important being the event record, represented
by the \texttt{Event} class. Actually, there are two objects of this
class, one called \texttt{process}, that only covers the few partons
of the hard process above, and another called \texttt{event}, that 
covers the full story from the incoming beams to the final hadrons. 
A small \texttt{Info} class keeps track of useful one-of-a-kind information, 
such as kinematical variables of the hard process.

There are also two incoming \texttt{BeamParticle}s, that keep track
of the partonic content left in the beams after a number of 
interactions and initial-state radiations, and rescales parton
distributions accordingly. 

The process library, as well as parametrisations of total, elastic 
and diffractive cross sections, are used both by the hard-process
selection machinery and the MI one. 

The \texttt{Settings} database keeps track of all integer, double,
boolean and string variables that can be changed by the user to steer
the performance of \textsc{Pythia}, except that 
\texttt{ParticleDataTable} is its own separate database.

Finally, a number of utilities can be used just about anywhere,
for Lorentz four-vectors, random numbers, jet finding, simple 
histograms, and for a number of other ``minor'' tasks.

\section{Program usage}

When you want to use \textsc{Pythia}~8 you are expected to provide 
the main program. At least the following commands should them be used:
\begin{itemize}
\item \texttt{\#include "Pythia.h"} to gain access to all the 
relevant classes and methods,
\item \texttt{using namespace Pythia8;} to simplify typing,
\item \texttt{Pythia pythia;} to create an instance of the generator,
\item \texttt{pythia.readString("command");} (repeated as required) to 
modify the default behaviour of the generator (see further below), or 
alternatively
\item \texttt{pythia.readFile("filename");} to read in a whole file of 
commands, one per line,
\item \texttt{pythia.init();} to initialize the generator, with different
optional arguments to be used to set incoming beam particles and energies,
\item \texttt{pythia.next();} to generate the next event, so this call
would be placed inside the main event generation loop,
\item \texttt{pythia.statistics();} to write out some summary information
at the end of the run.
\end{itemize}

The \texttt{pythia.readString(...)} and \texttt{pythia.readFile(...)} 
methods are used to modify the values stored in the databases,
and it is these that in turn govern the behaviour of the program. 
There are two main databases. 
\begin{itemize}
\item \texttt{Settings} come in four kinds, boolean \texttt{flag}s, 
integer \texttt{mode}s, double-precision \texttt{parm}s, and string 
\texttt{word}s. In each case a change requires a statement of the form 
\texttt{task:property = value}, e.g.\ \texttt{TimeShower:pTmin = 1.0}.
\item \texttt{ParticleDataTable} stores particle properties and decay 
tables. To change the former requires a statement of the form 
\texttt{id:property = value}, where \texttt{id} is the identity code 
of the particle, an integer. The latter instead requires the form 
\texttt{id:channel:property = value}, where \texttt{channel} is a 
consecutive numbering of the decay channels of a particle.
\end{itemize}
Commands to the two databases can be freely mixed. The structure with
strings to be interpreted also allows some special tricks, like that one
can write \texttt{on} instead of \texttt{true} and \texttt{off} instead of 
\texttt{false}, or that the matching to variable names in the databases is 
case-insensitive.

Information about all settings and particle data can be found in the online
manual, which exists in three copies. The \texttt{xml} one is the master
copy, which is read in when an instance of the generator is created, to set
up the default values that subsequently can be modified. The same information 
is then also provided in a copy translated to more readable \texttt{html}
format, and another copy in \texttt{php} format. The interactivity of the 
latter format allows a primitive graphical user interface, where a file of 
commands can be constructed by simple clicking and filling-in of boxes. 

The online manual contains more than 60 interlinked webpages, from a
program overview to some reference material, and in between extensive 
descriptions how to set up run tasks, how to study the output, and 
how to link to other programs. In particular, all possible settings 
are fully explained. 

\section{Trying it out}

If you want to try out \textsc{Pythia}~8, here is how:
\begin{itemize}
\item Download \texttt{pythia8108.tgz} (or whatever is the current 
version when you read this) from\\
\texttt{http://www.thep.lu.se/}$\sim$\texttt{torbjorn/Pythia.html}
\item \texttt{tar xvfz pythia8108.tgz} to unzip and expand.
\item \texttt{cd pythia8108} to move to the new directory.
\item \texttt{./configure ...} is only needed to link to external 
libraries, or to use options for debug or shared libraries, so can be
skipped in the first round.
\item \texttt{make} will compile in $1 - 3$ minutes (for an archive library, 
same amount extra for a shared one).
\item The \texttt{htmldoc/pythia8100.pdf} file contains A Brief Introduction
\cite{Sjostrand:2007gs}.
\item Open \texttt{htmldoc/Welcome.html} in a web browser for the full manual.
\item Install the \texttt{phpdoc/} directory on a webserver and open
\texttt{phpdoc/Welcome.php} in a web browser for an interactive manual.
\item The \texttt{examples} subdirectory contains $>30$ sample main programs:
standalone, link to libraries, semi-internal processes, \ldots  
\item These can be run by \texttt{make mainNN} followed by 
\texttt{./mainNN.exe > outfile}.
\item A \texttt{Worksheet} contains step-by-step instructions
and exercises how to write and run main programs.
\end{itemize}

Note that \textsc{Pythia} is constructed so it can be run standalone, 
and this is the best way to learn how it works. For an experimental
collaboration it would only be a piece in a larger software puzzle, 
and so a number of hooks has been prepared to allow various kinds of 
interfacing. The price to pay for using them is a more complex structure, 
where e.g.\ the origin of any errors is less easy to hunt down. Several 
aspects, such as the access to settings and particle data, should remain
essentially unchanged, however.   

\section{Outlook}

\textsc{Pythia}~6.4 is still maintained, with a current version 6.418
that weighs in at over 77,000 lines of code (including comments and
blanks) and has a 580 page manual \cite{Sjostrand:2006za}, plus update
notes and sample main programs. No further major upgrades will occur
with this program, however, and we intend to let it gradually die.

Instead \textsc{Pythia}~8.1 should be taking over. Currently it is smaller 
than its predecessor, with ``only'' 53,000 lines of code and a puny 27 page 
manual \cite{Sjostrand:2007gs}, but with much further online documentation 
and a big selection of sample main programs. It already contains several
features not found in 6.4,  and will gradually become the obvious version
to use. 

The LHC collaborations are strongly encouraged to accelerate the transition
from 6.4 to 8.1, e.g.\ by serious tests with small production runs, 
to find any remaining flaws and limitations.

\section*{Acknowledgements}

The \textsc{Pythia}~8 program was made possible by a three-year 
``sabbatical'' with the SFT group at CERN/EP. This unwavering support 
is gratefully acknowledged. Mikhail Kirsanov and other members of the 
GENSER group has provided further help with Makefiles and other technical 
tasks. The work was supported in part by the European Union Marie Curie 
Research Training Network MCnet under contract MRTN-CT-2006-035606. 

\bibliographystyle{heralhc} 
{\raggedright
\bibliography{lutp0817}
}
\end{document}